\documentclass[epsfig, 12pt, onecolumn]{article}

\usepackage{graphicx}

\newcommand{\centeron}[2]{{\setbox0=\hbox{#1}\setbox1=\hbox{#2}\ifdim 
                             \wd1>\wd0\kern.5\wd1\kern-.5\wd0\fi \copy0 
                             \kern-.5\wd0\kern-.5\wd1\copy1\ifdim\wd0>\wd1 
                             \kern.5\wd0\kern-.5\wd1\fi}} 
\newcommand{\ltap}{\>\centeron{\raise.35ex\hbox{$<$}} 
                     {\lower.65ex\hbox{$\sim$}}\>} 
\newcommand{\gtap}{\>\centeron{\raise.35ex\hbox{$>$}} 
                     {\lower.65ex\hbox{$\sim$}}\>} 

\newcommand{\newc}{\newcommand} 
\newc{\gsim}{\lower.7ex\hbox{$\;\stackrel{\textstyle>}{\sim}\;$}} 
\newc{\lsim}{\lower.7ex\hbox{$\;\stackrel{\textstyle<}{\sim}\;$}} 
\newc{\gev}{\,{\rm GeV}} 
\newc{\mev}{\,{\rm MeV}} 
\newc{\ev}{\,{\rm eV}} 
\newc{\kev}{\,{\rm keV}} 
\newc{\tev}{\,{\rm TeV}}

\newc{\mz}{M_Z} 
\newc{\mw}{m_{\rm weak}} 
\newc{\nr}[1]{N^c_R{}_{#1}} 
 
\def\beq{\begin{equation}} 
\def\eeq{\end{equation}} 
\def\bea{\begin{eqnarray}} 
\def\eea{\end{eqnarray}} 
\def\bitem{\begin{itemize}} 
\def\eitem{\end{itemize}} 
\newc{\ie}{{\it i.e.}}          \newc{\etal}{{\it et al.}} 
\newc{\eg}{{\it e.g.}}          \newc{\etc}{{\it etc.}} 
\newc{\cf}{{\it c.f.}} 
\def\bar#1{\overline{#1}} 
\def\vev#1{\left\langle #1 \right\rangle}

\def\inv{^{\raise.15ex\hbox{${\scriptscriptstyle -}$}\kern-.05em 1}} 
\def\lbar{{\lower.35ex\hbox{$\mathchar'26$}\mkern-10mu\lambda}} 

\let\<=\langle 
\let\>=\rangle

\let\+=\uparrow 

\def\singlespaced{\baselineskip=\normalbaselineskip}

\begin{document} 
\thispagestyle{empty} 
\vspace*{.5cm} 
\noindent 
\hspace*{\fill}{\large OUTP-04/17P}\\ 
\vspace*{2.0cm} 
 
\begin{center} 
{\Large\bf Asymmetric Sneutrino Dark Matter \\ and the 
$\Omega_{\rm{b}}/\Omega_{\rm{DM}}$ Puzzle} 
\\[2.5cm] 
{\large Dan Hooper$^1$, John March-Russell$^2$, and 
Stephen M. West$^2$}\\[.5cm] 
{\it $^1$Astrophysics Dept., University of Oxford,\\  
Denys Wilkinson Building, Oxford OX1 3RH, UK} 
\\[.2cm] 
{\it $^2$Rudolf Peierls Centre for Theoretical Physics,\\ 
University of Oxford, 1 Keble Road, Oxford OX1 3NP, UK} 
\\[.2cm] 
(October, 2004) 
\\[1.1cm]

{\bf Abstract}\end{center} 
\noindent 
The inferred values of the cosmological baryon and dark matter  
densities are strikingly similar, but in most theories of the early universe 
there is no true explanation of this fact; in particular, the baryon 
asymmetry and thus density depends upon unknown, and {\it a priori} 
unknown and possibly small, CP-violating phases which are independent of all parameters  
determining the dark matter density.  We consider models of dark matter 
possessing a particle-antiparticle asymmetry where this 
asymmetry determines both 
the baryon asymmetry and strongly effects the dark matter density, thus 
naturally linking $\Omega_{\rm{b}}$ and $\Omega_{\rm{dm}}$.  We show that sneutrinos 
can play the role of such dark matter in a previously studied variant 
of the MSSM in which the light neutrino masses result from 
higher-dimensional supersymmetry-breaking terms.

\newpage 
 
\setcounter{page}{1} 
 
\section{Introduction} 
 
For some time it has been apparent that the inferred values of the 
cosmological baryon and dark matter densities are strikingly similar. 
For example, the latest WMAP-determined range for the dark matter density, 
$0.129 > \Omega_{\rm{CDM}} h^2 > 0.095$, is within a factor of a few 
of the combined WMAP and big-bang nucleosynthesis determined value 
of the baryon density, $0.025 > \Omega_{\rm{b}} h^2 > 0.012$ 
\cite{WMAP,BBN}. 
In the vast majority of models of the early universe, the cosmological 
baryon and dark matter densities are independently determined. The 
surviving baryon density is set by a baryon asymmetry generated  
during baryogenesis, and thus depends upon unknown, and {\it a priori} 
unknown and possibly small CP-violating phases, as well as unknown baryon-number violating 
dynamics. In contrast, the dark matter density is set by the `freeze-out' 
of the interactions that keep the dark matter in equilibrium, and 
is independent of the dynamics of baryogenesis.   
In particular, 
although weakly-interacting massive particles at the TeV-scale  
naturally have a relic density of $\cal{O}$(1) times the critical 
density, this is not at all the case for baryons.  The Boltzmann 
equations show the size of the baryon relic density 
in the absence of an asymmetry is ${\cal O}(10^{-11})$. 
Thus the comparative 
closeness of the baryon and dark matter densities poses a puzzle. 
One possible approach to this problem is to more closely integrate 
the dynamics of baryogenesis with that of the origin of dark matter. 
In particular, it is natural to consider models where the dark matter 
and baryon sectors share a quantum number, either continuous or discrete, 
which provides a relation between their surviving number densities and thus 
energy densities.\footnote{An alternate approach is to invoke a form 
of the Anthropic Principle, see, for example,
Ref. \cite{wilczek}. Other ideas are explored in Refs. \cite{Kus,Enq,Oak2,Oak,farrar}.} 
 
Specifically, in this letter we consider models of dark matter 
possessing a particle-antiparticle asymmetry where this 
asymmetry strongly effects the dark matter density, and 
through the electroweak (EW) anomaly, determines the baryon asymmetry, thus 
naturally linking $\Omega_{\rm b}$ and $\Omega_{\rm CDM}$.  
(A noteworthy early attempt along these lines which shares some 
features with our model is contained in Ref.~\cite{kaplan}.) 
The shared quantum number clearly requires that the dark matter particle 
not be it's own antiparticle, so the putative dark matter candidate 
cannot be the standard LSP neutralino of the broken Minimal 
Supersymmetric Standard Model (MSSM).  In fact, 
we will show in sections~3 and 4 that sneutrinos can play the role 
of such dark matter in a previously studied variant 
of the MSSM in which the light neutrino masses result from 
higher-dimensional supersymmetry-breaking terms~\cite{ahhmsw,borz,sw, 
mrw,hmrw}.  This model preserves all the successes of the MSSM, such  
as stability of the weak scale and unification of gauge couplings,  
while being, at least in part, testable at the LHC.  
 
 
Before we proceed to the calculation of the relic dark 
matter and baryon densities and the details of our model, 
it is useful to consider some aspects of the 
idea of a shared quantum number determining the ratio of baryon 
to dark matter.  First, consider the simplified case 
in which, to a very good approximation, the dark matter and baryon 
sectors cannot exchange their conserved quantum number after 
its first production.  As an example, consider the situation 
in which the asymptotic baryon number $B=1$ states of the Standard 
Model (SM) have global charge, $q$, while the lightest asymptotic states 
(with mass $m_{\rm dm}$) in a hidden sector carry charge, $Q$.  Then 
conservation of global charge implies that $q\left(n_{\rm b} 
- n_{\rm \bar{b}}\right) =-Q\left( n_{\rm dm} - n_{\rm \bar{dm}} 
\right)$, where the $n$'s are the number densities of the 
indicated (anti)particles. 
Further assume that interactions in both sectors 
are strong enough such that almost all antiparticles are eliminated 
by annihilation with their particles (assuming $Q/q<0$ for simplicity). 
This implies 
\beq 
n_{\rm b} =  c n_{\rm dm} ~~{\rm with}~~ c = |Q/q| , 
\label{naive} 
\eeq 
which in turn leads to 
\beq 
\frac{\Omega_{\rm b}}{\Omega_{\rm dm}} = \frac{m_{\rm b}}{m_{\rm dm}} 
\frac{n_{\rm b}}{n_{\rm dm}} = c \frac{m_{\rm b}}{m_{\rm dm}} . 
\label{naive2} 
\eeq 
Thus the energy densities are indeed related, but the ratio is 
only naturally ${\cal O}(1)$ if the ratio of the baryon to dark 
matter masses is not too small.  It is important to note that 
this disfavours models of the above type where the dark matter 
candidate is a `hidden sector' particle 
and favours a particular class of models where the dark matter 
candidate has weak scale mass, its mass arising either 
from electroweak symmetry breaking, 
or from the dynamics which drives electroweak breaking, such as 
supersymmetry softly broken at the weak-scale.\footnote{ 
In extensions of the SM such as the MSSM where many 
gauge-non-neutral fermionic and scalar states also gain mass 
at this scale, the beta-function coefficient for 
$SU(3)_{\rm QCD}$ sharply increases as energies are reduced 
below the weak scale, and the confinement scale and thus baryon 
mass is naturally not very much smaller than $m_{\rm weak}$.} 
 
As a consequence, the naive statement of Eq.(\ref{naive}) also 
requires modification 
as the approximation of negligible interactions between the two 
sectors, and thus negligible exchange of the shared conserved 
quantum number, does not hold in realistic models.  In this  
situation, the ratio of the conserved quantum number stored 
in the two sectors, $c$, is not simply given by $Q/q$, but is determined 
by the `chemical' equilibrium conditions between the two sectors 
just before the freeze-out of the relevant interactions.  We will 
further discuss this point in section 2.  Moreover, two other 
alterations to the naive relation Eq.(\ref{naive2}) can be 
present.  First, the shared quantum number may only be 
conserved modulo $n$, in which case `self' annihilations may occur, 
and second, it is not always the case that dark matter interactions are 
efficient enough such that $n_{\rm \bar{dm}} \ll n_{\rm dm}$.   
Nevertheless, as we argue in the next section, a careful 
consideration of such models shows that a version of the 
naive statement Eq.(\ref{naive2}) can indeed hold.

\section{The Effect of a Matter-Antimatter Asymmetry on the  
Density of a Thermal Relic} 
\label{reliccalc} 
 
If there were an asymmetry between the density of dark matter particles 
and dark anti-matter particles in the early universe, it may have a 
considerable effect on the relic density calculation \cite{griestasy}. In this section, 
we review the relic density calculation for a single species, two 
species with no asymmetry and two species with an asymmetry. 
 
If dark matter consists of a single species, the relic density 
calculation is straightforward. The most common example of a single 
species dark matter candidate would be a neutralino LSP with 
negligible coannihilations with other SUSY particles. In such a case, 
the resulting relic density is simply 
\begin{equation} 
\Omega h^2 \cong \frac{1.04 \times 10^9}{M_{\rm{Pl}}} 
\frac{x_{F}}{\sqrt{g_*}} \frac{1}{(a+3b/x_F)}, 
\label{onesp} 
\end{equation} 
where $M_{\rm{Pl}}$ is the Planck mass, $x_F$ is the WIMP's mass over 
the freeze-out temperature, $g_*$ is number of relativistic degrees of 
freedom available at freeze-out and $a$ and $b$ are related to the 
WIMP's self-annihilation cross section by $\sigma v = a + b v^2 + 
\mathcal{O}$$(v^4)$, $v$ being the relative velocity of the 
annihilating WIMPs. For a weakly interacting relic, $x_F$ is 
generally about 20 and varies only mildly with changes in the cross 
section or mass. 
 
If we instead consider a dark matter candidate {\em and} its antiparticle, 
the calculation becomes somewhat different. Labeling these species 1 
and 2, we now have each particle's self-annihilation cross section, 
$\sigma_{11}=\sigma_{22}$, and the particle-antiparticle annihilation 
cross section, $\sigma_{12}=\sigma_{21}$. In the limit that 
$\sigma_{12}$ is much smaller than $\sigma_{11}$, we simply get two 
decoupled species which each provide the relic density of 
Eq.(\ref{onesp}). Otherwise, the effective cross section becomes 
$\sigma_{11}+\sigma_{12}$ for each species. For the two species 
together, a density of 
\begin{equation} 
\Omega h^2 \cong 2 \times \frac{1.04 \times 10^9}{M_{\rm{Pl}}} 
\frac{x_{F}}{\sqrt{g_*}} 
\frac{1}{(a_{11}+a_{12}+3(b_{11}+b_{12})/x_F)} 
\label{twosp} 
\end{equation} 
is produced. This case is essentially the calculation of the density 
of a relic with the coannihilation of a degenerate particle \cite{griestthree}. 
 
If an asymmetry exists between the number density of particles and 
antiparticles in the early universe, again this calculation becomes 
modified. The most well known example of this is the baryon-antibaryon 
asymmetry which leads to the current baryon density of the universe 
despite the very large baryon-antibaryon annihilation cross 
section. In such a case, where the particle-particle and 
antiparticle-antiparticle annihilation cross sections ($\sigma_{11}$ 
and $\sigma_{22}$) are zero, or negligibly small, the minimum relic 
density can be related to the magnitude of the particle-antiparticle 
asymmetry: 
\begin{equation} 
\Omega h^2_{\rm{min}} = \Omega h^2_{\rm{bary}} 
\frac{A}{A_{\rm{bary}}} \frac{m}{m_{\rm{bary}}}, 
\label{estimate} 
\end{equation} 
where $\Omega h^2_{\rm{bary}}$ is the current density of baryons in 
the universe, $A$ and $A_{\rm{bary}}$ are the particle-antiparticle 
asymmetries of our relic and of baryons, defined by $A=(n - \bar{n}) 
/n$. $A_{\rm{bary}}$ is known to be of the order of $10^{-9}$. $m$ and 
$m_{\rm{bary}}$ are the masses of our relic and of baryons ({\it i.e.} 
the proton mass). Of course, if the particle-particle annihilation 
cross section for the relic is not negligible, Eq.(\ref{estimate}) will 
not hold. There will be a generic tendency for the density of a relic 
to move towards this value as a result of an asymmetry, however. 
 
To accurately assess the effect of an asymmetry on the thermal relic 
abundance, the effective cross sections must be determined by 
integrating over the thermal history. In particular, for each species 
separately, the densities predicted are 
\begin{equation} 
\Omega_{1} h^2 \cong \frac{1.04 \times 10^9}{M_{\rm{Pl}}} 
\frac{x_{F}}{\sqrt{g_*}} 
\frac{1}{(a_{11,\rm{eff}}+a_{12,\rm{eff}}+3(b_{11,\rm{eff}}+b_{12,\rm{eff}})/x_F)} 
\label{oneasy} 
\end{equation} 
and 
\begin{equation} 
\Omega_{2} h^2 \cong \frac{1.04 \times 10^9}{M_{\rm{Pl}}} 
\frac{x_{F}}{\sqrt{g_*}} 
\frac{1}{(a_{22,\rm{eff}}+a_{21,\rm{eff}}+3(b_{22,\rm{eff}}+b_{21,\rm{eff}})/x_F)}, 
\label{twoasy} 
\end{equation} 
where the effective cross sections are given by 
\beq 
\begin{array}{rclrcl} 
a_{11, \rm{eff}} &=& a_{11},\qquad  & a_{12, \rm{eff}} &=& x_F \int^{\infty}_{x_F} 
\frac{a_{12}}{x^2} \bigg(\frac{e^{-x}}{e^{-x}+A}\bigg) dx, \nonumber\\ 
 a_{22, \rm{eff}} &=& a_{22}, \qquad & a_{21, \rm{eff}} 
&=& x_F \int^{\infty}_{x_F} \frac{a_{12}}{x^2} 
\bigg(\frac{e^{-x}+A}{e^{-x}}\bigg) dx, \nonumber\\ 
b_{11, \rm{eff}} &=& b_{11}, \qquad & b_{12, \rm{eff}} &=& 2 x_F^2 
\int^{\infty}_{x_F} \frac{b_{12}}{x^3} 
\bigg(\frac{e^{-x}}{e^{-x}+A}\bigg) dx, \nonumber\\ 
b_{22, \rm{eff}} &=& b_{22}, \qquad & b_{21, 
\rm{eff}} &=& 2 x_F^2 \int^{\infty}_{x_F} \frac{b_{12}}{x^3} 
\bigg(\frac{e^{-x}+A}{e^{-x}}\bigg) dx . 
\end{array} 
\eeq 
The factor, $e^{-x}/(e^{-x}+A)$, is the depletion of the second 
species due to the asymmetry at a temperature, $T=m/x$. The effect of 
this depletion depends on the size of the asymmetry and the cross 
sections involved. If the estimate of Eq.(\ref{estimate}) is 
considerably less than the standard two-species result of 
Eq.(\ref{twosp}), then the asymmetry has little impact and the sum of 
Eqs.(\ref{oneasy}) and (\ref{twoasy}) is nearly equal to 
Eq.(\ref{twosp}). If, however, the estimate of Eq.(\ref{estimate}) is 
larger than the standard two-species result, the relic density moves 
from the value found with Eq.(\ref{twosp}) towards that of 
Eq.(\ref{estimate}). If $\sigma_{11}/\sigma_{12}$ is small, then the 
result will be closer to Eq.(\ref{estimate}). If 
$\sigma_{11}/\sigma_{12}$ is large, then the density will be less 
modified. Note that in some cases, the effect of an asymmetry can be 
to lower the density of a relic by up to a factor of two. This occurs 
when $\sigma_{11}$ and $\sigma_{12}$ are comparable thus efficiently 
deplete the second species while leaving the first largely 
uneffected. 
 
Although these effects are rather model dependent, the net effect can be large regions of parameter space which predict the observed 
density of dark matter for a WIMP with a mass near $m \sim m_{p} \, 
\Omega_{\rm{WMAP}} h^2/\Omega_{\rm{bary}} h^2$, for a 
matter-antimatter asymmetry similar to that for baryons. For 
concreteness, we consider a specific example of such a model in the 
following section.

\section{An Example: Mixed Sneutrino Dark Matter} 
 
Within the context of the Minimal Supersymmetric Standard Model 
(MSSM), sneutrinos do not make a very appealing dark matter 
candidate. They annihilate quite efficiently, 
resulting in a relic density smaller than the observed dark matter 
density for sneutrinos lighter than about 500 GeV. Furthermore, their 
elastic scattering cross section is sufficiently large to be easily 
observed by direct dark matter experiments.  Thus it is natural 
that in contrast to other 
supersymmetric dark matter candidates, sneutrino dark matter has 
received considerably less attention \cite{sneutrinodarkmatter}. 
 
Looking beyond the vanilla MSSM, however, it is quite easy to construct a 
viable model for sneutrino dark matter. For example, sneutrinos 
could mix through $A$-terms with a singlet forming a mixed sneutrino 
state. This would reduce the annihilation cross section, potentially 
providing the appropriate quantity of dark matter, as well as reduce 
the elastic scattering cross section to experimentally acceptable 
levels.  In fact, such a scenario has previously been discussed as a way 
to reconcile the results of the direct dark matter detection 
experiments DAMA and CDMS \cite{sw}.  These models 
have light sneutrino mass eigenstates which are mixtures of 
left-handed `active' sneutrinos, and right-handed `sterile' 
sneutrinos. 
 
Such mixing naturally arises in models in which the light neutrino 
masses are generated by supersymmetry breaking effects 
\cite{ahhmsw,borz,mrw,hmrw}.  The importance to us of these models is that 
the light sneutrino states share a non-anomalous $(B-L)$-symmetry 
with the baryons which is only weakly broken by the Majorana 
neutrino masses. In analogy to the Giudice-Masiero 
mechanism for the Higgs $\mu$-term there exists a global symmetry 
that forbids a direct mass, $M_R N N$, and neutrino Yukawa coupling, 
$\lambda L N H_u$, involving the SM singlet right-handed neutrino superfield 
$N$, but allows non-renormalizable couplings to a 
supersymmetry-breaking hidden-sector field $T$. 
The right-handed neutrino mass and the neutrino Yukawa coupling, $\lambda L N 
H_u$, then only arise from $1/M_{\rm planck}$-suppressed terms 
involving the fundamental supersymmetry breaking scale 
$m_I \sim ({v M_{pl}})^{1/2}$, where $v$ is the weak scale. 
The relevant terms in the Lagrangian are 
\beq 
\mathcal{L} = \int d^4 \theta \left( h \frac{T^\dagger}{M} N N + 
{\tilde h} \frac{T^\dagger T}{M^2} N^\dagger N + h_B \frac{T^\dagger 
TT^{\dagger} }{M^3} N N + \ldots \right) + 
\int d^2\theta \left( g\frac{T}{M} L N H_u  \right),  
\label{newL} 
\eeq 
where all dimensionless couplings $g,h$, etc, are taken to be 
${\cal O}(1)$. The structure of the Lagrangian can be enforced by 
 some simple symmetries \cite{ahhmsw,borz,mrw}, including R-parity,
 which leads to a stable lightest supersymmetric particle
 (LSP) dark matter candidate.
After supersymmetry is broken in the hidden 
sector at the scale $m_I$, the $F$ and $A$ components of the  
field $T$ acquire intermediate scale vevs, $\vev{F_T} \sim \vev{A_T}^2 
\sim m_I^2$.  Substituting these vevs into  
the Lagrangian shows that after SUSY breaking, there arises: 
(1) a neutrino Yukawa coupling of size $\sim 10^{-7}-10^{-8}$, 
(2) right-handed neutrino and sneutrino masses at the {\em weak scale}, 
(3) a weak-scale trilinear scalar $A$-term, and 
(4) a right-handed sneutrino lepton-number violating $B$-term 
with magnitude $B^2 \sim ({\rm few}\times 100\mev)^2$.  
These together lead to two sources of neutrino masses, a tree level 
see-saw contribution, as well as a dominant 1-loop contribution, 
leading to a consistent and attractive neutrino phenomenology 
\cite{ahhmsw,mrw}. 
 
More relevant for us, the sneutrino mass matrix squared is 
\beq 
\left(\begin{array}{cc} 
\tilde{\nu}^* & n \end{array}\right)\left 
(\begin{array}{cc}m_L^2 & \frac{1}{\sqrt{2}} Av\sin\beta  \\ 
\frac{1}{\sqrt{2}} Av\sin\beta & m_R^2 \end{array} \right) 
\left(\begin{array}{c} \tilde{\nu} \\ 
n^*\end{array}\right), 
\label{snumassmatrix} 
\eeq 
where we have ignored the small lepton number violating $B$-term for 
the moment.  Therefore mixing between the left-handed `active' sneutrino, 
$\tilde{\nu}$, and the right-handed `sterile' sneutrino, $\tilde{n}$, occurs, 
producing the light mass eigenstate $\tilde{\nu}_1=-\tilde{\nu}\sin\theta 
+ \tilde{n}^* \cos\theta$, 
where $\tan 2\theta = - \sqrt{2} Av \sin\beta/(m_R^2 - m_L^2)$, 
with $m_R^2 = M_N^2 + m^2_{R,{\rm soft}}$ and 
$m_L^2 = m_Z^2\cos 2\beta /2 + m^2_{L,{\rm soft}}$.   
Since the coupling of the lighter sneutrino eigenstate to the $Z$ is 
suppressed by $\sin \theta$, the direct LEP experimental 
constraints are weakened, and the calculation of the sneutrino 
relic abundances is modified.  A numerical analysis of the susy 
parameter space shows that a mixing $\sin\theta \simeq 0.3$ 
is quite natural.\footnote{Note that the size of the $A$ terms 
is limited by considerations of vacuum stability \cite{hmrw}.} 
(The $B^2$ mass-term splits the CP-even and CP-odd components  
of $\tilde{\nu}_1$ by the amount $\delta \sim \cos^2\theta B^2/m_1$. 
This can have important consequences for detection, potentially  
making consistent the DAMA and CDMS experiments depending upon 
parameters \cite{sw}.  For the purposes of our discussion 
this splitting is irrelevant.)

We now turn to the calculation of the relative asymmetry in the 
sneutrino and baryon sectors.  The method is a simple adaption 
of the standard `chemical' equilibriation techniques applied 
to, for example, the calculation of the ratio $B/(B-L)$ in the 
SM~\cite{chemical} or MSSM~\cite{iims} in the presence of
anomaly-induced baryon number violating processes 
in the early universe. 
For each species of particle we introduce a chemical potential, 
$\mu_A$, and for each interaction in equilibrium, say $A+B 
\leftrightarrow C +D$ there is a corresponding relation among 
chemical potentials $\mu_A +\mu_B = \mu_C + \mu_D$.   
Note that when in {\em thermal} (not necessarily chemical) 
equilibrium the CP-conjugate species $\bar{A}$ has 
$\mu_{\bar A} = - \mu_A$.  The asymmetry in particle-antiparticle 
number density for the relevant case of $\mu\ll T$ and a 
species of mass $m$ is given by  
$n - {\bar n} = g T^2 \mu/6$ for $T\gg m$, and is Boltzmann 
suppressed by $\exp(-m/T)$ for $T\lsim m$. 
 
The electroweak $B+L$ anomaly 
interactions provide an efficient mechanism to 
transmute between $B$- and $L$-number densities, at least above the EW phase 
transition temperature $T_c$ where they are rapid.  Our task is  
to calculate the ratio of sneutrino to baryon asymmetries 
at $T_c$ where the exchange between the two sectors 
switches off.  (We implicitly assume that the EWPT is 1st order 
so that anomaly-induced interactions drop out of equilibrium below 
$T_c$.  For realistic values of $T_c$ the results of our 
calculation are not particularly 
sensitive to this assumption.)  Above the EWPT 
temperature the chemical potentials for all SM gauge bosons vanish, 
so, for example, members of an EW doublet have the same $\mu$. 
Moreover, efficient flavour mixing among quarks induced by 
EW interactions such as $u_{Li} + \phi^0_1 \leftrightarrow u_{Rj}$, 
involving the neutral up-like higgs, or $d_{Li} + \phi^0_2 
\leftrightarrow d_{Rj}$, involving the neutral down-like higgs, ensures 
that only one chemical potential, say that for $u_L$, is 
required to describe the quark sector.   
Similarly, efficient flavour mixing among the light neutrinos 
holds for mixing angles, $\theta_{ij} \gsim 10^{-7}$, 
so that only one chemical potential, $\mu_{\nu_L}$, is needed to
describe the neutrino 
and sectors.  In addition non-perturbative EW processes can be thought 
of creating, say, a neutron made of lhd quarks, and a lhd neutrino 
out of the vacuum for each generation. When this process is in equilibrium it 
imposes an additional relation which can be written  
\beq 
3\mu_{u_L} + \mu_{\nu_L} = 0 . 
\label{nonpert} 
\eeq 
 
When the MSSM susy spectrum is included the analysis becomes 
 more involved. To simplify the analysis let 
us first assume at a temperature $T$ (with 
$T>T_c$) that the MSSM susy spectrum, including $k$ 
rhd sneutrinos can be considered light ($m\lsim T$). As 
shown in Ref.\cite{iims} the usual 
soft-susy-breaking terms and supersymmetric partners of the  
SM interactions lead to enough relations among the chemical 
potentials that only 3 are needed for an independent set:  
$\mu_0 \equiv (\mu_{\phi_{10}} - \mu_{\phi_{20}})/2$ in the neutral 
Higgs sector, $\mu_{u_L}$, and $\mu_{\nu_L}$.  In our case 
we must add an additional $k$ light ($m<T_c$) rhd 
neutrinos and sneutrinos with their associated potentials $\mu_{\nu_{Ri}}$ and 
$\mu_{\tilde \nu_{Ri}}$, but we also have $A$-term interactions 
between the lhd and rhd sneutrinos, and the very small 
lepton-number violating rhd sneutrino $B$-term.  These  
together enforce the conditions 
\beq 
\mu_{\tilde{\nu_R}} = \mu_{\tilde{\nu_L}} = \mu_{{\nu_L}}, 
\label{nuchemical} 
\eeq 
where the first equality arises from $A$-term mediated 
interactions (as the $A$ terms are ${\cal O}(m_w)$ it is simple 
to show that these are in equilibrium), and the second equality 
arises from $SU(2)_w\times U(1)_Y$ gaugino 
interactions.\footnote{Note that the rhd neutrinos only interact  
via the small neutrino Yukawa, $\lambda \sim 10^{-7}-10^{-8}$, so 
these states are not in equilibrium at $T\sim T_c$.  However, 
$\mu_{\nu_R}$ does not enter our final expression 
for the sneutrino to baryon number density ratio, so its 
value is irrelevant.}  In terms of  
$\mu_0, \mu_{u_L}$, and $\mu_{\nu_L}$, the condition 
Eq.(\ref{nonpert}) is unchanged in the case of the 
${\rm MSSM}+k$ rhd neutrino supermultiplets. 
 
The final step in the calculation of the relative number 
densities proceeds by expressing the relevant densities in terms of 
the independent chemical potentials.  For the baryon number 
density $B$ the result is simply 
\beq 
B = 6T^2 \mu_{u_L} . 
\label{bdensity} 
\eeq 
For the sneutrino asymmetry the result depends upon the mass 
spectrum of the sneutrino mass eigenstates relative to $T_c$. 
In the range of parameters of interest to us where the light sneutrino 
eigenstate $\tilde{\nu}_1$ is the LSP we have $m_1 < T_c$. 
On the other hand the heavier eigenstate $\tilde{\nu}_2$ 
can have $m_2 > T_c$ or $m_2 < T_c$.  If both eigenstates satisfy $m<T_c$ 
then the sneutrino density is (allowing for a possible $k$ such states) 
\beq 
S = \frac{k T^2}{3} (\mu_{\tilde \nu_L} + \mu_{\tilde \nu_R}) = 
\frac{2k T^2}{3} \mu_{\nu_L} 
\label{snudensity} 
\eeq 
while if only $m_1<T_c$ then $S$ is half as large.   Finally 
using the non-perturbative relation Eq.(\ref{nonpert}) we find 
that 
\beq 
\frac{A}{A_{\rm bary}} = -\frac{S}{B} = \frac{k}{3}~~{\rm to} ~~\frac{k}{6} 
\label{asymresult} 
\eeq 
depending on $m_2$. We view $k=1$ as most likely, so we 
specialize to that case in the next section. Finally we mention 
that a full susy-mass-threshold dependent calculation of the relative 
asymmetries introduces only small corrections to this result (see 
Ref.\cite{iims} for a similar analysis in the MSSM case). 
 
In the end the vital point is that independent of whatever 
dynamics at scales 
$E>T_c$ produces an asymmetry in either the baryon sector, or in the 
neutrino or sneutrino sector, the $(B+L)$-anomaly-induced interactions 
together with EW gaugino and $A$-term interactions 
automatically distributes the asymmetry between the baryons and 
the light {\it sneutrino} states, with a predictable $A/A_{\rm{bary}}$ 
ratio.  The asymmetry 
could originate from, e.g. a GUT-based baryogenesis mechanism, or maybe 
more interestingly in the context of the sneutrino dark matter 
model there is the possibility that it could arise via a 
resonant leptogenesis mechanism at the TeV-scale as 
discussed in Ref.\cite{hmrw}.  The end result is that 
we expect $1/3 \gsim A/A_{\rm{bary}} \gsim 1/6$ independent 
of the source of the asymmetry.

\section{Asymmetric sneutrino and baryon abundances} 
 
We will now apply the formalism of section~\ref{reliccalc} to the 
sneutrino model.  
Given the value of $1/3 \gsim A/A_{\rm{bary}} \gsim 1/6$, we expect the 
role of the matter-antimatter asymmetry to be most important for 
sneutrinos with masses on the order of $m \sim {\rm few} \times m_{p} \, 
\Omega_{\rm{WMAP}} h^2/\Omega_{\rm{bary}} h^2 \simeq 30$ GeV. For this 
reason, we will focus on the mass range of about 10 to 80 GeV. In this 
range, the most important annihilation channel is s-channel 
$Z$-exchange to fermions. This is a p-wave amplitude, and has no 
$a$-term in the expansion in $v^2$ (see section~\ref{reliccalc}). This 
is also only a sneutrino-antisneutrino process ({\it i.e.} 
$\sigma_{11, 22}$, not $\sigma_{12, 21}$). Other channels include 
t-channel neutralino and chargino exchanges to neutrinos and charged 
leptons. The channel to charged leptons includes both $a$ and $b$ 
terms in the $v^2$ expansion. Additionally, sneutrino-sneutrino pairs 
(or antisneutrino-antisneutrino pairs) can annihilate via t-channel 
neutralino exchange to neutrino pairs of the same flavor. This is the 
only contribution to $\sigma_{12, 21}$ in this model. 
 
The annihilation cross sections for the neutralino and chargino 
exchange processes depend on the masses and mixings of these exchanged 
particles, although the $Z$-exchange channels do not. We have adopted 
the parameters $M_1$=300 GeV, $M_2$=300 GeV, $\mu$=600 GeV and $\tan 
\beta=50$ throughout our calculation.  Additionally, there is the 
mixing angle between the sneutrino and the singlet which can be 
varied. This quantity has the effect of reducing all electroweak 
couplings of the sneutrino eigenstate by a factor of $\sin^2 \theta$, 
{\it i.e.} reducing the annihilation cross sections by a factor of 
$\sin^4 \theta$. 
 
Finally, one can consider the s-channel exchange of Higgs bosons to $b 
\bar{b}$ (or $ZZ$ and $W^+W^-$ above the appropriate mass 
thresholds). This process is not suppressed by $\sin^4 \theta$. The 
efficiency of this process depends on the magnitude of the trilinear 
scalar coupling and is only important near the Higgs resonance 
($m_{\tilde{\nu}} \simeq m_{h}/2$) or above the gauge boson mass 
thresholds. 
 
\begin{figure}[tb] 
\centering\leavevmode 
$\begin{array}{c@{\hspace{0.5in}}c} 
\includegraphics[width=0.32\textwidth]{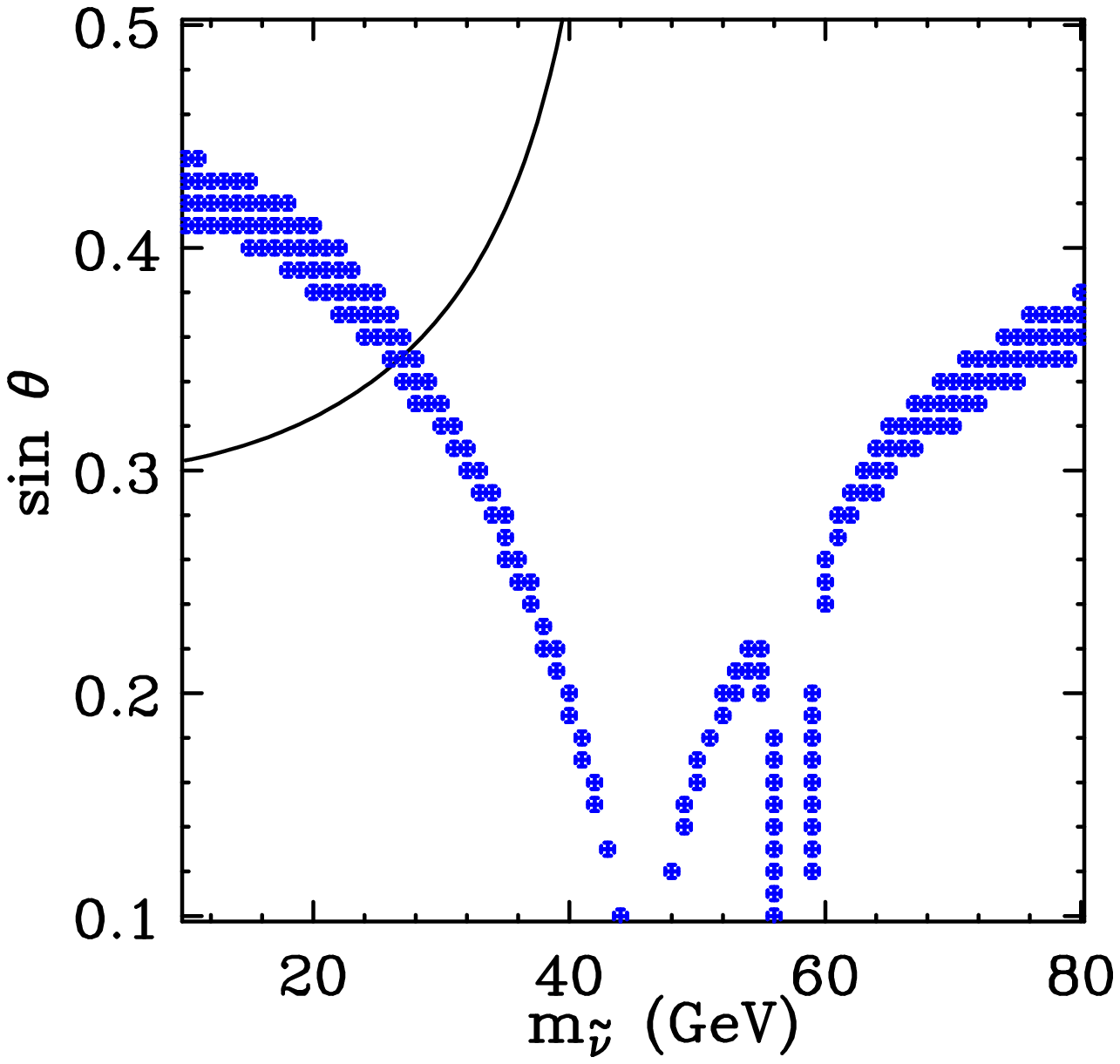}  
\includegraphics[width=0.32\textwidth]{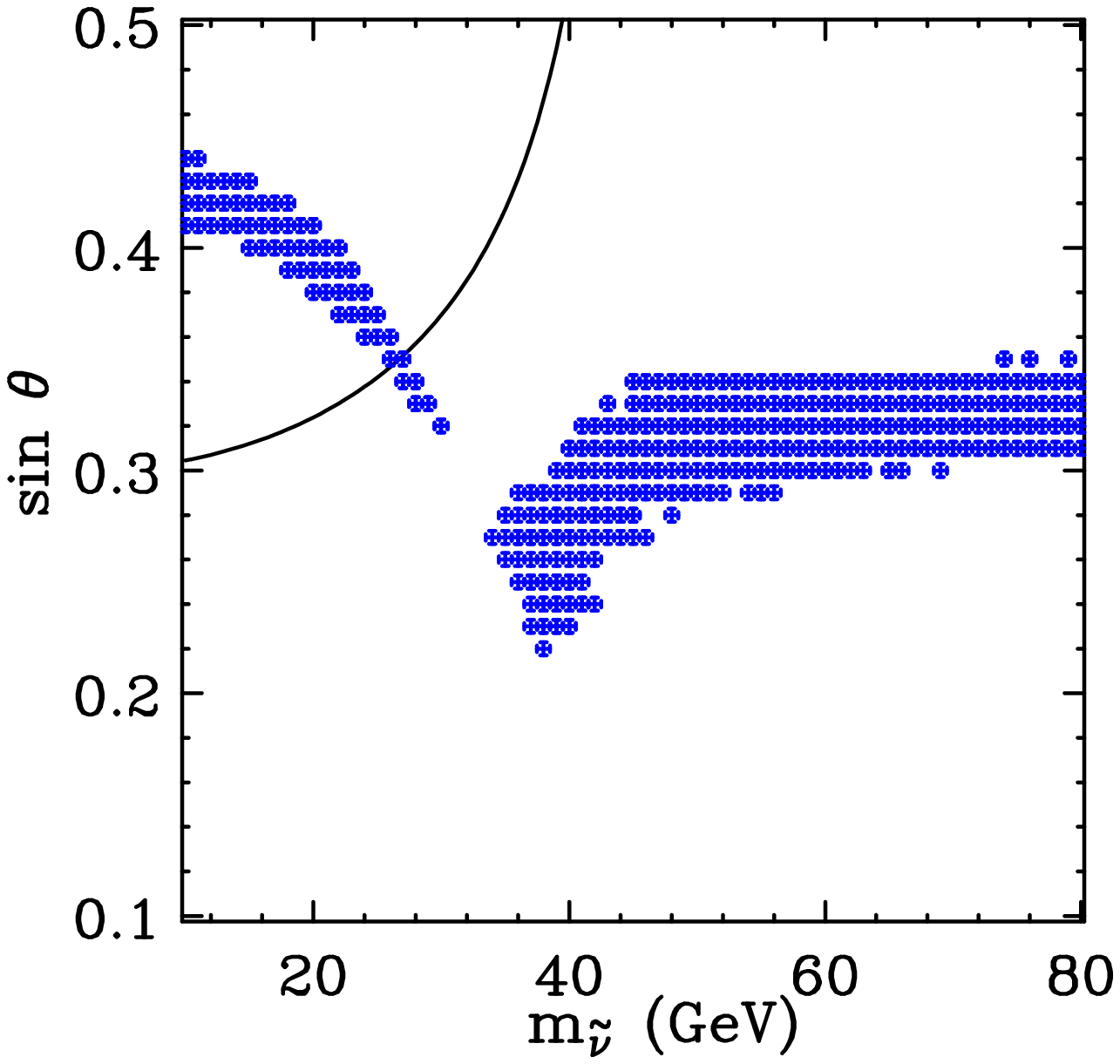}  
\includegraphics[width=0.32\textwidth]{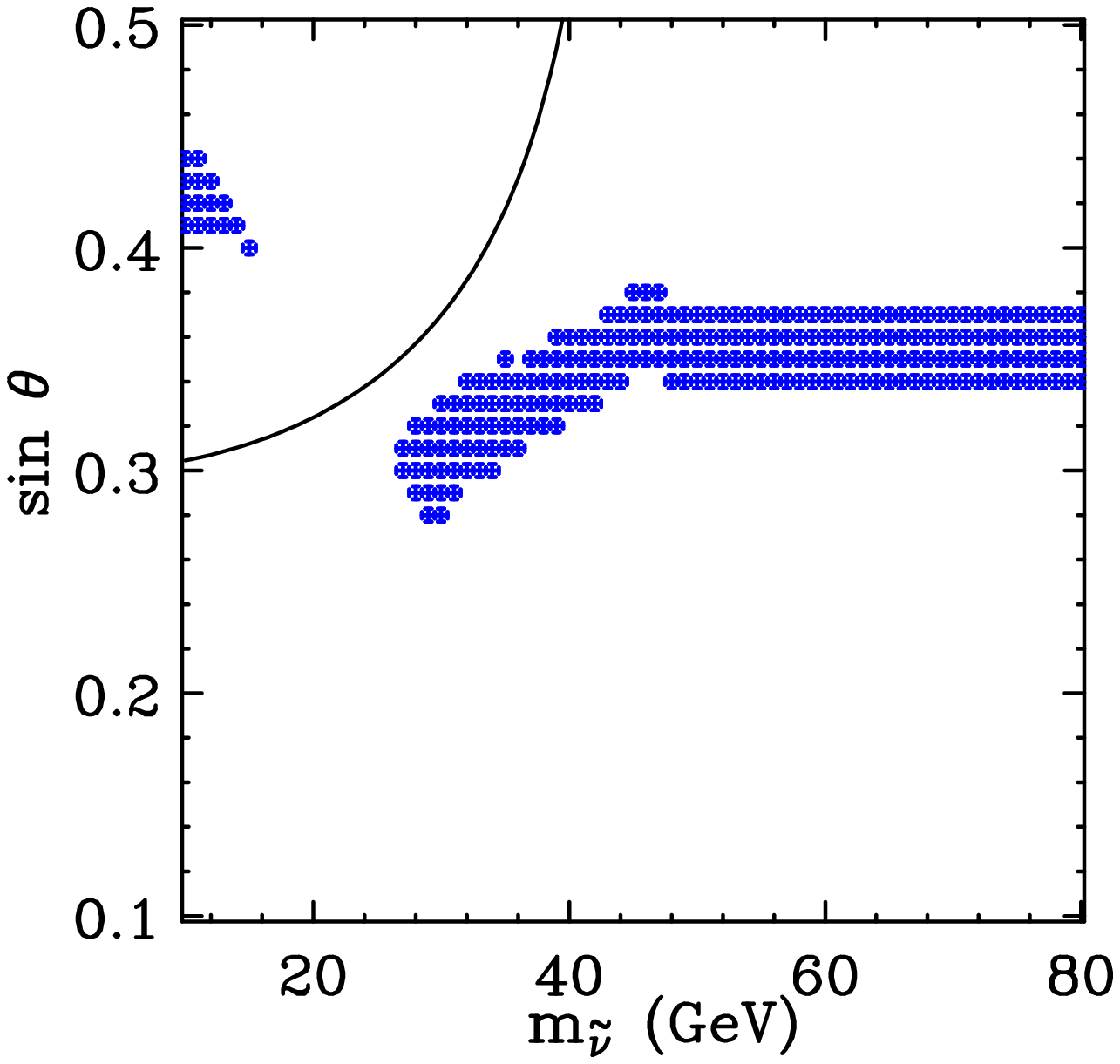}\\ [0.4cm] 
\end{array}$ 
\caption{The regions of parameter space which provide the quantity of 
mixed sneutrino cold dark matter measured by WMAP, $0.129 > 
\Omega_{\rm{CDM}} h^2 > 0.095$. In the left frame, the standard 
calculation with no matter-antimatter asymmetry is used. In the center 
frame the effect of a matter-antimatter asymmetry with 
$A/A_{\rm{bary}} \simeq 1/6$ is included, while in the right frame 
a matter-antimatter asymmetry $A/A_{\rm{bary}} \simeq 1/3$ is assumed. 
Notice the much larger regions of acceptable parameter space in cases 
with asymmetry. We use the following parameters: $M_1$=300 GeV, $M_2$=300 GeV, 
$\mu$=600 GeV, $\tan \beta=50$ and $m_h$=115 GeV. The region above the 
solid line is excluded by measurements of the invisible $Z$ decay 
width \cite{invisiblez}.} 
\label{relicplot} 
\end{figure}

\begin{figure}[tb] 
\centering\leavevmode 
\includegraphics[width=0.35\textwidth]{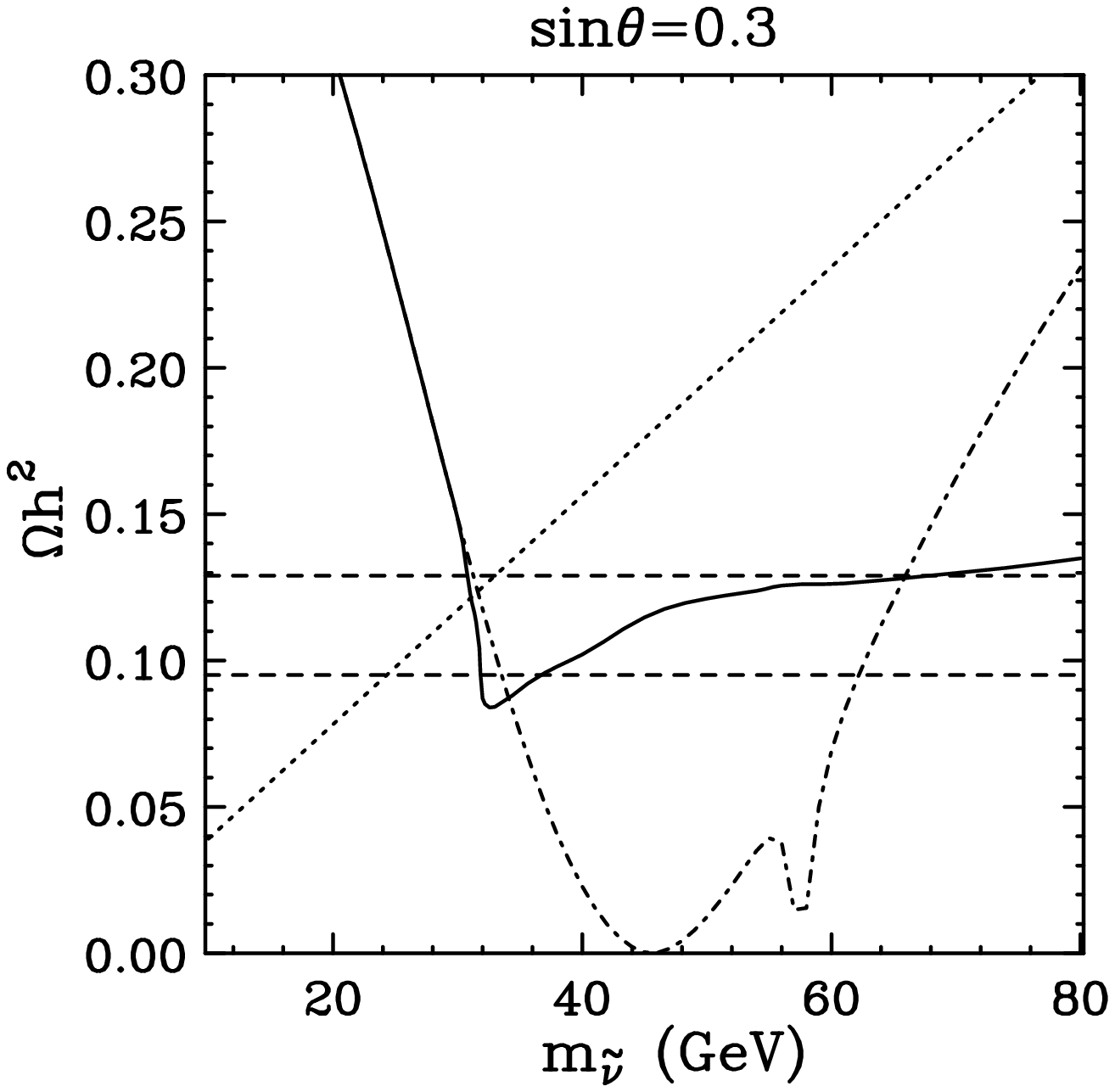}  
\caption{The thermal relic density as a function of mass for 
sneutrinos+antisneutrinos with no asymmetry (dot-dash), with a 
matter-antimatter asymmetry of $A/A_{\rm{bary}} \simeq 1/6$ (solid) 
and the estimate of Eq.(\ref{estimate}) (dots). The relic density range 
favored by WMAP is bound by dashed lines ($0.129 > \Omega_{\rm{CDM}} 
h^2 > 0.095$). $\sin \theta$=0.3, $M_1$=300 GeV, $M_2$=300 GeV, 
$\mu$=600 GeV, $\tan \beta=50$ and $m_h$=115 GeV have been used. See 
text for details.} 
\label{cutplot} 
\end{figure}

Our results are shown in figure~\ref{relicplot}. The shaded regions of 
the parameter space predict a relic density within the range measured 
by WMAP ($0.129 > \Omega_{\rm{CDM}} h^2 > 0.095$). In the left frame, 
no matter-antimatter asymmetry was included. This frame can be easily 
understood as requiring greater suppression of electroweak couplings 
(smaller $\sin \theta$) as the sneutrino mass approaches the $Z$ 
mass. The dip at 56-59 GeV is due to s-channel higgs exchange to $b 
\bar{b}$. In the right frame, a matter-antimatter asymmetry of 
$A/A_{\rm{bary}} \simeq 1/6$ was included. To further illustrate this 
effect, we show the result of this calculation across one value of 
$\sin \theta$ in figure~\ref{cutplot}. Below about 30 GeV, the 
matter-antimatter asymmetry has little effect on the calculation and 
the solid and dot-dashed lines fall nearly on top of each other. In 
the range of roughly 30-70 GeV, however, the asymmetry pulls the relic 
density above the standard symmetric result into the range favored by 
WMAP. Above this range, sneutrino-antisneutrino annihilation 
decreases, leading to larger relic densities for the case with no 
asymmetry. The relic density for the asymmetric case, however, is 
largely determined by the sneutrino-sneutrino annihilation cross 
section in this region, so does not increase as rapidly, therefore
resulting in a relic density much closer to the preferred value, even
for $m_{\tilde{\nu}}>70$~GeV.

\section{Indirect Detection} 
 
Sneutrinos in the mass range we have studied here annihilate 
dominantly through s-channel $Z$ exchange in the early universe. This 
annihilation channel consists of a $b$-term in the expansion $\sigma v 
= a + b v^2 + \mathcal{O}$$(v^4)$, and thus is very inefficient for 
annihilations at low velocities which take place in the late 
universe. At low velocities, the process $\tilde{\nu} \tilde{\nu} 
\rightarrow \nu \nu$, via t-channel neutralino exchange, is the most 
important annihilation channel. 
 
Given these characteristics, searches for gamma-rays, anti-protons, 
positrons or other particles produced in sneutrino annihilations in 
the Galactic halo are quite inefficient. Due to the rather large 
elastic scattering cross sections of sneutrinos, however, indirect 
detection techniques involving the capture of dark matter in bodies 
such as the Sun can be very effective. 
 
Neutrino telescopes detect neutrinos produced in dark matter 
annihilations in the core of the Sun by observing 'tracks' of muons 
generated in charged current interactions. If our sneutirno
\footnote{In the model presented, the dark matter candidate is
  actually an antisneutrino and consequently antineutrinos will be
  produced by dark matter annihilations in the sun.}  dark 
matter candidate is of muon or tau flavor, the flux of muons generated 
in this way in a detector at Earth is of the order of $10^{-12}$ 
cm$^{-2}$ s$^{-1}$ \cite{sw}, in conflict with data from the 
Super-Kamiokande and MACRO experiments, for example~\cite{macro}. If 
the candidate, however, is of electron flavor, these bounds 
can be easily averted. 
 
\section{Conclusions} 
 
In the standard freeze-out calculation for a weakly interacting dark  
matter relic, there is little reason to expect a density of dark matter  
which is similar to the density of baryons, short of anthropic arguments.  
In this letter, however, we have presented an explanation for the  
similarity between these two quantities. Our solution introduces an  
asymmetry between dark matter particles and anti-particles which is  
related to the baryon-antibaryon asymmetry. This leads to a natural dark  
matter relic density of the same order of magnitude as the baryon density. 
 
As an example, we considered a mixed sneutrino dark matter candidate which  
transfers its particle-antiparticle asymmetry to the baryons through 
the electroweak  
anomaly. We carry out the relic density calculation for such a candidate  
and find substantial regions of parameter space in which the observations  
of WMAP can be satisfied. With no asymmetry, only a narrow strip of  
parameter space can provide the observed relic density. 
 
\vskip 0.05in
\begin{center}
{\bf Acknowledgments}
\end{center}
\vskip0.05in
We wish to thank Yuval Grossman for a useful correspondence. SW is supported
by PPARC Studentship Award PPA/S/S/2002/03530.

\end{document}